\documentclass[%
preprint,
 amsmath,amssymb,
 aps, physrev,
]{revtex4-2}

\usepackage{graphicx}
\usepackage{dcolumn}
\usepackage{bm}
\usepackage[english]{babel}

\begin{document}


\title{\textbf{Polarization memory effect in a multimode fiber} 
}%

\author{Gauri Arora}
\affiliation{%
 Advanced Research Center for Nanolithography (ARCNL), Science Park 106, Amsterdam, 1098 XG, The Netherlands
}%

\author{Lyubov V. Amitonova}%
 \email{Contact author: l.amitonova@vu.nl}
\affiliation{
 Advanced Research Center for Nanolithography (ARCNL), Science Park 106, Amsterdam, 1098 XG, The Netherlands
}%
\affiliation{
Department of Physics and Astronomy, Vrije Universiteit Amsterdam, De Boelelaan 1105, 1081HV Amsterdam, The Netherlands
}%

\date{\today}

\begin{abstract}
Optical memory effects are well-known types of amplitude-domain wave correlation enabling control over light scattered through diffusive materials or multimode fibers. In this letter, we report the phenomenon of random polarization memory effect. We observe that, despite strong scattering, which leads to random spatial polarization distributions at the multimode fiber output, local polarization states of the optical field maintain a high degree of correlation with respect to the input polarization state. This newly observed effect reveals an additional layer of deterministic behavior in light propagation through multimode fibers and opens up new opportunities for imaging and sensing.
\end{abstract}

\maketitle



Multimode fibers (MMF) are crucial tools for various optical applications~\cite{cao_controlling_2023, amitonova_multimode_2024}, including imaging \cite{Abdulaziz2023, doi:10.1126/science.abl3771, Amitonova:18}, sensing \cite{8733026}, metrology \cite{10.1063/5.0089159}, and communication \cite{10.1063/5.0049022}, to name a few. 
The light beam undergoes randomization in amplitude, phase, and polarization due to complex mode structure, mode coupling, modal dispersion, and polarization effects. This scrambling of light through MMFs limits their direct practical applications in imaging and other fields.  Several solutions have been reported to mitigate these effects of scrambling and precisely control the amplitude, phase, and polarization through an MMF \cite{cao_controlling_2023}. These techniques usually utilize wavefront shaping at the input, which is based on a scalar or vector transmission matrix-based approach \cite{popoff_measuring_2010, 10.1063/1.5136334}. 

Interestingly, despite strong scattering, optical fields often preserve a degree of spatial correlation, which has recently enabled a new class of imaging methods through highly scattering tissues. These new methods exploit what is traditionally called \emph{optical memory effect}~\cite{freund_memory_1988, feng_correlations_1988, osnabrugge_generalized_2017}.
In the context of MMFs, hidden correlations between the seemingly random output field distribution and the input beam have been extensively studied: rotational~\cite{Amitonova:15}, axial \cite{10.1063/5.0067892} and translational \cite{Caravaca-Aguirre:21} memory effects have been demonstrated.
These memory effects describe how specific transformations of the input wavefront produce correlated changes in the output field, despite the complex modal mixing within the fiber and have potential for imaging applications \cite{Li2021, doi:10.1073/pnas.2221407120, PhysRevLett.61.2328}. However, they all describe amplitude-related transformations only -- the hidden correlations between input and output polarization distributions have never been explored.

Meanwhile, the vectorial nature of electromagnetic waves plays an indispensable role in imaging and light-matter interactions. Control over the polarization state of light is important in single-molecule detection, optical tweezers, and optical coherence tomography~\cite{cao_controlling_2023}. A particular state of polarization will be scrambled when transmitting through an MMF, as the modes generally do not maintain a linear polarization state due to spin-orbit interaction~\cite{1073624, Rochat:03}. When light is coupled into an individual mode, it will spread to other modes, each of which will experience a distinct polarization change. Moreover, fiber deformation, residual birefringence, or strain lead to polarization scrambling. Polarization dynamics with wavelength, space, and time has been studied in a few-mode fiber~\cite{Fridman:12}. The coupling between the spatial and polarization degrees of freedom in an MMF opens the possibility of utilizing the spatial wavefront shaping of the input wave to control the polarization state of the MMF output~\cite{ploschner_seeing_2015, xiong_complete_2018, mitchell_high-speed_2016, Mounaix2019}. However, this polarization control requires polarization-resolved complex transmission matrix measurements, which directly link amplitude, phase and polarization state for the sets of input and output fields of the given MMF and it's particular configuration.


In this Letter, we report, for the first time, the predictable relationship between the input and output polarization states in an MMF with random polarization scrambling and an unknown transmission matrix. Specifically, we demonstrate that rotating the input polarization induces spatially resolved periodic changes in the intensity of the linear polarization components across the output speckle pattern. Remarkably, we observe that, despite the randomized structure, the polarization state at each spatial point at the MMF output rotates synchronously with the input polarization. We term this phenomenon the polarization memory effect of an MMF. This newly observed effect reveals an additional layer of deterministic behavior in light propagation through MMFs and similar scattering media and opens up new opportunities for imaging and sensing.

 \begin{figure*}[!htb]
  \includegraphics[trim=8mm 8mm 4mm 5mm, clip,
width=\columnwidth]{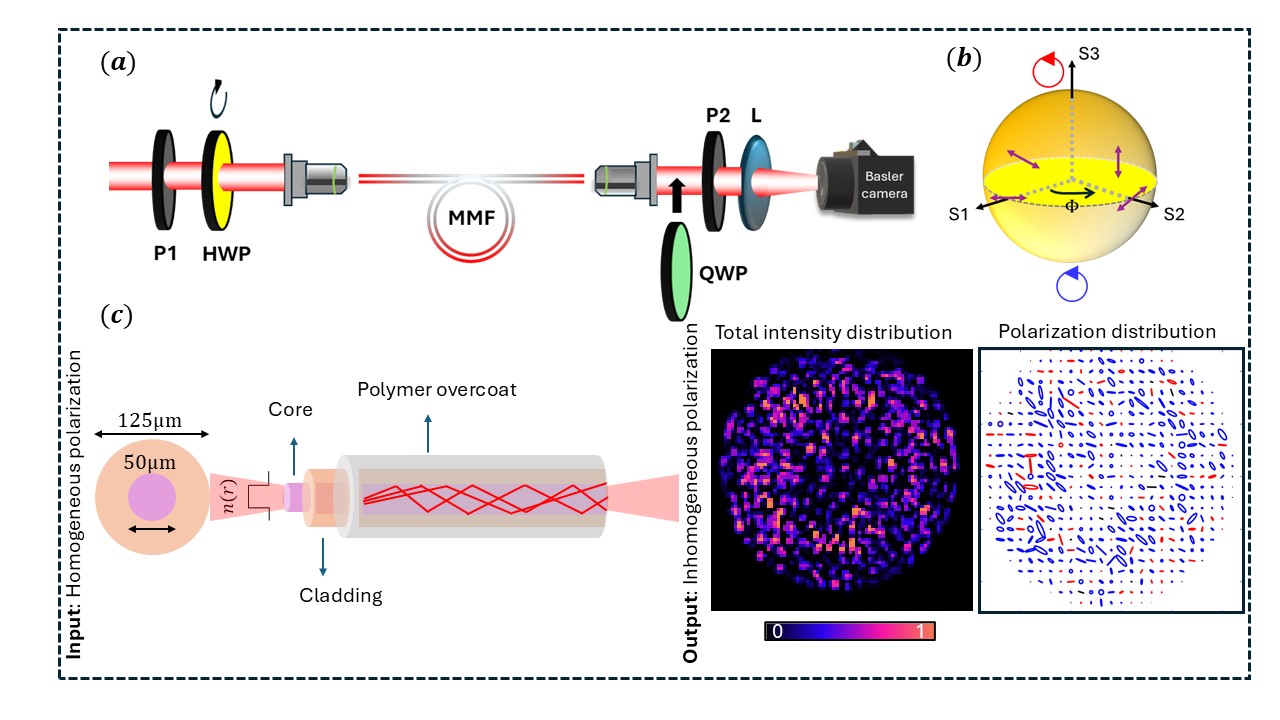}
 \caption{(a) Schematics of experimental setup. Coherent, cw, linearly polarized light is coupled into a multimode fiber by a microscope objective. The input polarization is controlled using a half-wave plate. Abbreviations: P1,2, polarizer; L, lens; H(Q)WP, half(quarter) wave plate; MMF, multimode fiber. (b) Depiction of the polarization states and Stokes parameters on Poincar\'e sphere. (c, left side) Sketch of a multimode fiber. (c, right side) Experimental results: the speckle intensity distribution at the multimode fiber output facet (left) and the corresponding polarization distribution (right). Red and blue colors in polarization distribution represent right and left-handed polarization states. The size of the polarization states is scaled as per the total intensity distribution. The input initial polarization state is oriented in a $-40$ degrees direction.
 \label{Fig. 1}}
  \end{figure*}


Our experimental setup is schematically shown in Fig. 1(a). A homogeneously linearly polarized collimated beam from HeNe laser is focused at the input facet of an MMF using a microscopic objective (Olympus, $40\times$, NA$_{eff}$ = 0.25). The beam size at the back focal plane of the objective is adjusted to match the NA of the fiber, thus minimizing coupling losses. In the experiments, a round core MMF (Thorlabs FG050LGA) with core diameter $50\mu m$, NA $0.22$ and length $50$ cm is used. The fiber was loosely coiled during measurements, as schematically shown in Fig. 1(a). The input facet of the MMF is depicted in the left side of Fig. 1(c).
A half-wave plate (HWP) after the polarizer P1 is used to rotate the azimuth of the input linear state of polarization. The output facet of the MMF is imaged using another microscopic objective and a tube lens on a Basler camera with a pixel size of $3.45~\mu$m. A  second polarizer (P2) is placed before the camera, and the output linear polarization component intensity distributions can be recorded with a controlled precise rotation of the HWP. A combination of a quarter-wave plate (QWP) and a polarizer (P2) are also used to perform Stokes polarimetry of the output speckle patterns for different rotations of the HWP.

To determine the spatially resolved polarization distribution at the output of the MMF, six intensity distributions corresponding to different polarization components are recorded for each specific HWP angle. These measurements enable the calculation of the four Stokes parameter distributions, which fully describe the polarization state at each spatial point. These Stokes parameters are further processed in MATLAB software to reconstruct the spatially resolved polarization map at the MMF output. Each polarization state can be represented mathematically as four measurable Stokes parameters \cite{goldstein_polarization_2010}. The spatially varying Stokes parameter can be mathematically expressed as:
\begin{equation}
\centering
\begin{aligned}
S_0(x,y)=|E_H(x,y)|^2+|E_V(x,y)|^2=I_H(x,y)+I_V(x,y),\\
S_1(x,y)=|E_H(x,y)|^2-|E_V(x,y)|^2=I_H(x,y)-I_V(x,y),\\
S_2(x,y)=2Re(E_H^{*}(x,y)\cdot E_V(x,y))=I_D(x,y)-I_A(x,y),\\
S_3(x,y)=2Im(E_H(x,y)\cdot E_V^{*}(x,y))=I_{R}(x,y)-I_{L}(x,y).
\end{aligned}
\end{equation}

Here, $H$, $V$, $D$, $A$, $R$, and $L$ represent horizontal, vertical, diagonal, anti-diagonal, right-circular, and left-circular polarization components, respectively. $S_0(x,y)$, $S_1(x,y)$, $S_2(x,y)$, and $S_3(x,y)$ are spatially varying Stokes parameters where $x,y$ represent spatial coordinates.

The experimental results for a single input field with linear polarization state are presented in Fig. 1(c), where the right side shows the total speckle intensity distribution and the corresponding polarization distribution. The red and blue colors in the polarization distribution correspond to two different handednesses. As expected, the scrambling of light through an MMF results in an inhomogeneously polarized speckle pattern and the information on the original polarization state appears to be lost in the random output field.

To confirm our experimental results, we also simulate light propagation through an MMF. We use a semi-analytical vectorial mode solver~\cite{snyder_optical_1983}. In short, we solve the wave equation in a cylindrical geometry with known boundary conditions and compute the (complex-valued) electromagnetic field components $E_{\phi,n}(x,y)$ and $E_{\rho,n}(x,y)$, and the (real-valued) propagation constant $\beta_n$ of all the $n$ modes. The excitation field is decomposed into the basis of fiber modes. Then, each mode propagates along the length of the fiber and consequently acquires a different phase, creating a seemingly random interference pattern.

\begin{figure}[t]
  \centering
  \includegraphics[trim=3mm 70mm 60mm 0mm, clip,
width=0.82\columnwidth]{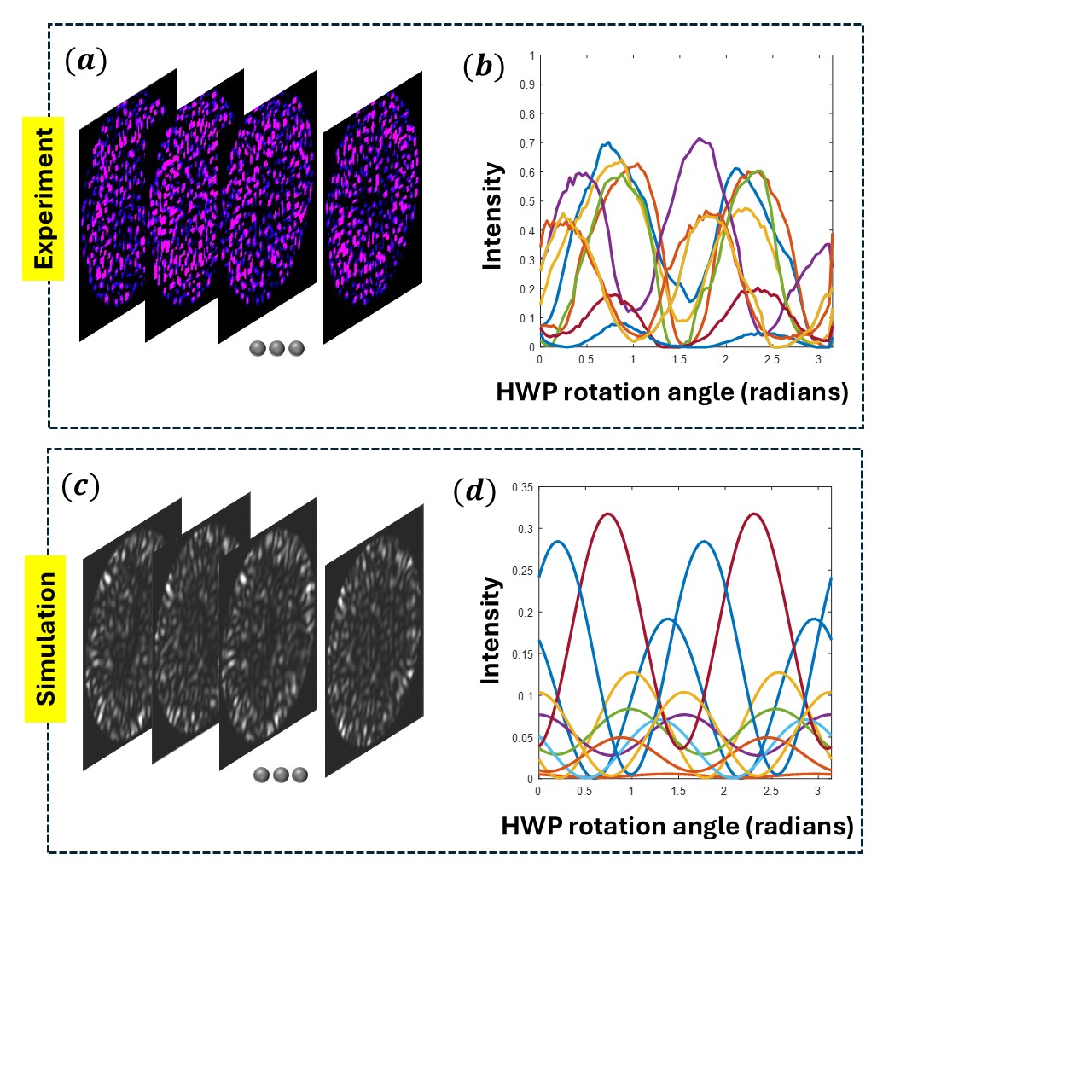}
 \caption{Manifestation of the polarization memory effect in a multimode fiber. The left side depicts (a) experimentally measured and (b) simulated speckle intensity distributions of a selected linear polarization component at the output for four different input polarization states. (b, c) Experimentally measured (b) and simulated (c) intensity variations at ten randomly selected speckle points at the MMF output facet as the input polarization is rotated using the half-wave plate. 
 \label{Fig. 2}}
  \end{figure}

In the first set of experiments, a second polarizer (P2) without the QWP is placed before the camera, and the output linear polarization component intensity distributions are recorded with a controlled rotation of the HWP at the MMF input from $0$ to $180$ degrees with a step size of $2$ degrees using a motorized stage.
The experimental results are presented in Fig. 2 (top).
Figure 2(a) shows the four-component speckle patterns at the MMF output for four different input polarization states. Ten random points in the speckle pattern are selected and the intensity variations averaged over $3 \times 3$ pixel region, which corresponds to $10.35 \times 10.35 \mu$m are plotted as a function of the HWP rotation angle. The results are shown in Fig. 2(b). We see that the local intensity at the MMF output varies periodically with the polarization angle at the MMF input. The Fast Fourier Transforms have been calculated and confirmed that the intensity fluctuation frequency corresponds to $90$ degrees of the HWP rotation. The simulation results shown in Fig. 2 (c,d) confirm the findings. 

In the second set of experiments, Stokes polarimetry is performed to record not only intensity variations but also the entire polarization distribution across the MMF output facet. Here a combination of a QWP and a polarizer (P2) is used and Stokes polarimetry is performed for different rotations of the HWP at the MMF input from $0$ to $90$ degrees (clockwise direction) in a step size of $5$ degrees.
The results are presented in Fig. 3. Three random points in the MMF output are selected and the polarization state is plotted with the rotation of the HWP angle, as shown in Fig. 3(a). It is observed that despite the random output polarization distribution across the MMF output, the polarization state of each local point rotates by the exact same amount as the input polarization.

The observed phenomena can be explained as following. Each polarization state in the spatial polarization distribution can be represented as a point on the Poincar\'e sphere constructed using normalized Stokes parameters. The rotation of input polarization state using HWP can be understood as the phase difference ($\phi$ shown in Fig. 1(b)) introduced between the right and left circular polarization components (geometrical phase) and hence the beam traversing on the latitude of the Poincar\'e sphere. It is observed that all the local polarization states at the output MMF facet rotate as per the azimuth of the input polarization state. This effect can be interpreted as the indirect transfer of phase-difference information between the input orthogonal circular polarization states to the output. This phase difference information is reflected in the output polarization distribution, while the total speckle intensity distribution remains the same. This is because the rotation of HWP maintains the same amplitude of orthogonal circular polarization states. In addition, rotation of the HWP only alters the polarization state without affecting the total intensity of the input beam, ensuring that the total power remains conserved. The information can be translated into the intensity distribution by introducing a polarizer after the fiber (as has been shown in the first set of measurements), which effectively converts a vector beam interference to a scalar beam interference. The effect of introducing a linear phase difference variation at the input or by rotating the HWP is seen as the periodic variation of polarization component speckle intensity at each point of the output speckle distribution.

\begin{figure*}[!ht]
  \centering
  \includegraphics[trim=15mm 35mm 25mm 0mm, clip,
width=\columnwidth]{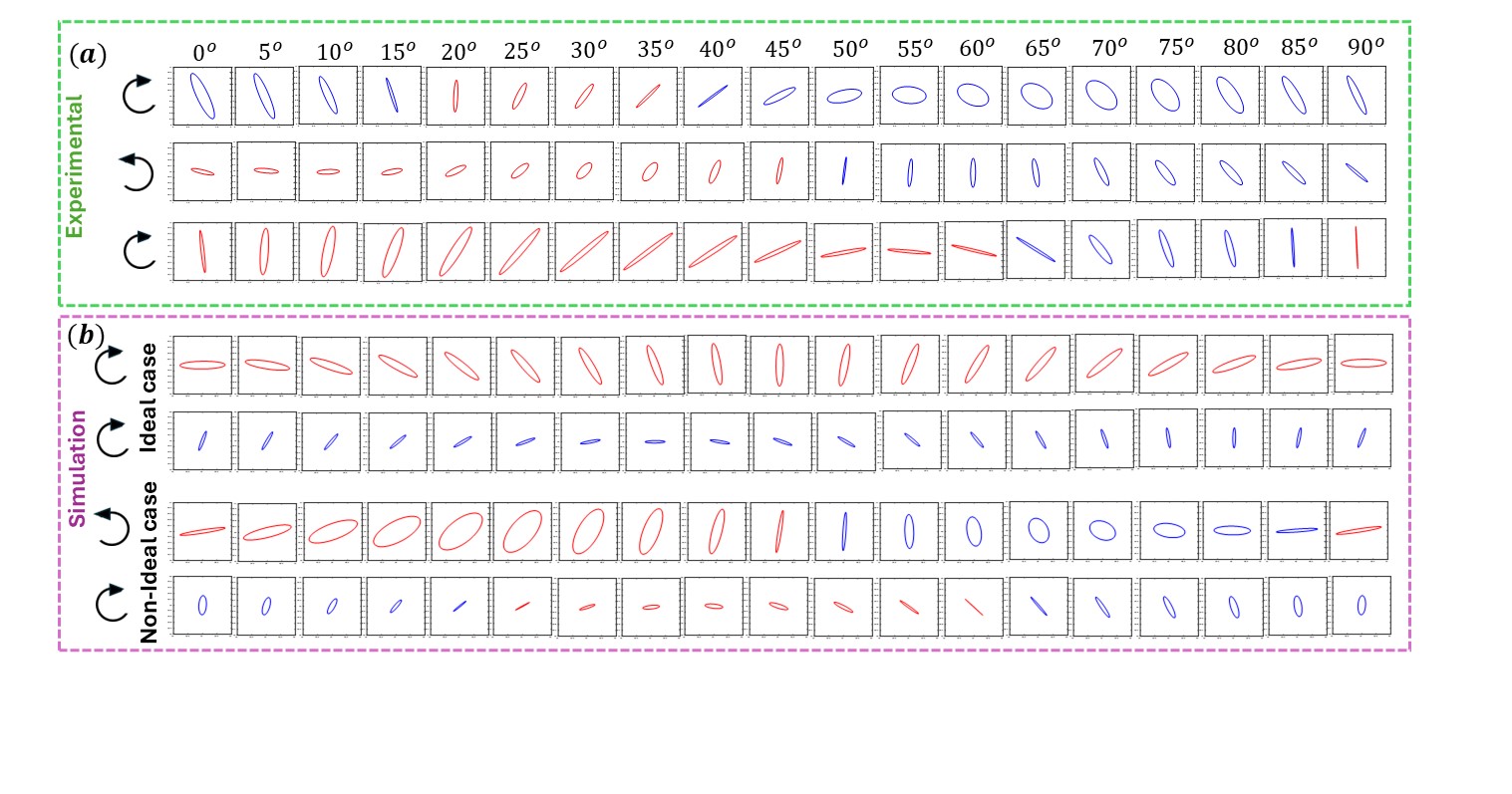}
 \caption{Random polarization memory effect in a multimode fiber. (a) Experimentally measured local polarization state variations at three randomly selected points at the MMF output as a function of the input half-wave plate rotation angle. (b) Simulated output polarization states at two randomly chosen points at the multimode fiber output as a function of the input half-wave plate rotation angle. Results are shown for both ideal (top) and non-ideal (bottom) conditions. The sizes of the polarization ellipses are scaled according to the total intensity at each point. 
 \label{Fig. 3}}
  \end{figure*}

Interestingly, as presented in Fig. 3(a), some of the points rotate in the clockwise direction while other points rotate in the anti-clockwise direction. This positive and negative phase difference ambiguity, which results in a different sense of rotation of different spatial polarization states, is attributed to the introduced birefringence in the MMF in experiments due to stress and the presence of a coil in the fiber. To confirm this, both the ideal case and non-ideal case with birefringence are simulated and studied.
The simulation results are presented in Fig. 3(b).  For a non-ideal case, a constant retardation of $\pi/3$ is introduced between the orthogonal linear polarization states. Two random points in the output polarization distribution are selected and the change in polarization state is plotted with respect to the input HWP rotation. In an ideal case, it is observed that each polarization state in the random output polarization distribution rotates in the same direction and by the same amount as the input polarization state without affecting the handedness and ellipticity of the polarization state. For the non-ideal case, there is a change in ellipticity and handedness of polarization states in addition to the rotation of azimuth with rotation of input polarization state. Also, it is observed that some of the polarization states rotate in the clockwise direction while others anti-clockwise. 

In addition, it is observed that due to the geometrical phase introduced by the rotating HWP at the input, the variation in polarization follows the polarization continuity rule \cite{Fösel_2017}. Every L-point crossing in experimental and simulated data (non-ideal case) with HWP rotation is accompanied by a handedness change of polarization state. This L-point or L-line singularity normally occurs in the spatial polarization distribution. However, we observed this polarization topological point in polarization variation with the rotation of the input polarization state using the HWP.

To summarize, we have demonstrated both theoretically and experimentally a previously unreported random polarization memory effect in an MMF. We show that the local polarization at each point in the output speckle pattern rotates in direct correspondence with the input polarization state. This predictable behavior persists despite strong light scrambling within the MMF, revealing an additional degree of deterministic behavior in scattered light propagation. The discovery adds a new dimension to the known family of memory effects. It opens promising opportunities for non-invasive imaging and could enhance signal stability and transmission efficiency in optical communication systems.

\section*{Data Availability Statement}
Data underlying the results presented in this paper are not publicly available at this time but may be obtained from the authors upon reasonable request.

\begin{acknowledgments}
This work was conducted at the Advanced Research Center for Nanolithography, a public-private partnership between the University of Amsterdam, Vrije Universiteit Amsterdam, University of Groningen, the Netherlands Organization for Scientific Research (NWO), and the semiconductor equipment manufacturer ASML, and was (partly) financed by ‘Toeslag voor Topconsortia voor Kennis en Innovatie (TKI)’ from the Dutch Ministry of Economic Affairs.
\end{acknowledgments}


\bibliography{apssamp}

\end{document}